# Jovian Zonal Winds Revealed from Cassini/VIMS Observations

**Shenghan Ma[1,2], Yuming Wang[1,2,3,*], Tao Li[1,2], Quanhao Zhang[1,2], Jiajia Liu[1,2] and Ru-obing Zheng[1,2]**

[1] National Key Laboratory of Deep Space Exploration/School of Earth and Space Sciences, University of Science and Technology of China, Hefei 230026, China

[2] CAS Center for Excellence in Comparative Planetology/CAS Key Laboratory of Geospace Environment/Mengcheng National Geophysical Observatory, University of Science and Technology of China, Hefei 230026, China

[3] Hefei National Laboratory, University of Science and Technology of China, Hefei 230088, China

*Corresponding author: Yuming Wang (ymwang@ustc.edu.cn)

## Key Points:

- A correlation-based method is developed for the Cassini/VIMS-IR observations to reveal zonal winds.
- Zonal winds are retrieved at several altitudes and latitudes, demonstrating their complexity with altitude.
- The vertical wind structure in the troposphere at the equator is obtained, showing eastward wind shear.

**Abstract**

Understanding Jupiter's zonal winds is crucial to unraveling the dynamics of its atmosphere. Over the last decades, multiple data sources and techniques have been used to study zonal winds in Jupiter. Here, we develop a correlation-based method for the near-infrared data from the Cassini spacecraft to investigate zonal winds at different altitudes. The new method uses Jupiter's rotation to scan the planet as it rotates, allowing to retrieve winds from the analysis of light-curves of specific pixels over the Jovian disc. The method allows to retrieve winds at multiple wavelengths from the Cassini/VIMS spectral data in spite of the low spatial resolution and the non-uniform cadence of the data. By applying this method to two VIMS data cubes acquired on January 15, 2001 at 09:42 UT and January 16, 2001 at 03:22 UT, we reveal the zonal winds at five main latitudes using information from three different wavebands, as well as the wind vertical structure at the equator, showing significant vertical wind shear in the troposphere. The vertical wind shear we derived is weaker than reported in previous studies, highlighting the intricate interactions among multiple dynamical processes in Jupiter's atmosphere and reflecting the complexity of its atmospheric circulation. Despite the uncertainty due to the low spatial/temporal resolution and non-uniform cadence of the Cassini/VIMS-IR spectral data, the new method established in this study maximizes the value of the Cassini/VIMS in understanding Jupiter's zonal winds. Further observations are essential to explore the underlying mechanisms in Jupiter's atmosphere.

**Plain Language Summary**

Jupiter's colorful striped appearance is related to its strong zonal winds, which shape the planet's atmosphere and play a significant role in the formation of the alternating bright zones and dark belts. Over the years, scientists have used data from multiple spacecraft to understand how these winds vary across different regions of Jupiter. In our work, we used data from the VIMS instrument on the Cassini mission and applied a new method to measure zonal winds in different altitudes in Jupiter's atmosphere. By comparing these winds at the equator, we learned more about how different parts of Jupiter's atmosphere interact.

## 1 Introduction

The gas giant of our solar system are particularly noted for their striking banded appearance. With the largest and most dynamic atmosphere, Jupiter stands out as the most prominent one. The banded appearance results from complex atmospheric dynamics and is characterized by alternating regions of dark and bright clouds. The latitudinal pattern of dark belts and bright zones correlates with the zonal winds (Ingersoll et al., 2004), and has been suggested to be related to secondary meridional circulations influenced by the internal heat and external solar irradiation (Fletcher et al., 2020).

In recent decades, our understanding of Jupiter's atmospheric behavior has advanced significantly, largely thanks to data provided by space missions. Zonal winds are a key element of the atmosphere (Ingersoll et al., 2004). Most previous studies focused on the zonal winds at the cloud top located at ~600 mbar based on the observations from, e.g., Voyager 1 and 2 (Limaye, 1986, 1989), Hubble Space Telescope (HST) Wide Field Camera 3 (WFC3) (Asay-Davis et al., 2011; García-Melendo, 2001; Tollefson et al., 2017), and Cassini Imaging Science Subsystem (ISS) (Porco et al., 2003). These studies been revealed that the zonal winds' distribution at the cloud top is not symmetrical. Each hemisphere of Jupiter has 6-7 eastward jets separated by westward jets. The strongest jet is eastward, located in the northern hemisphere, with a maximum measured velocity of 180 m/s at the cloud top (Limaye, 1989; Porco et al., 2003; Asay-Davis et al., 2011; Barrado-Izagirre et al., 2013; Tollefson et al., 2017; Johnson et al., 2018). At polar latitudes, the zonal jets weaken and the atmospheric circulation transitions into a regime dominated by polar vortices.

Notably, Li et al. (2006) used the ultraviolet images captured by Cassini/ISS to derive zonal winds above the cloud top. This was the first time zonal winds at ~350 mbar were directly measured, which increase slightly with altitude at the equator. Combining with Cassini CIRS thermal winds results (Flasar et al., 2004), they certified that the zonal winds are weaker with increasing altitudes except near zero latitude. They also derived zonal winds below the cloud top by tracking features within hot spots in continuum-band images. The zonal winds are suggested to be ~3 bar, compared with Galileo's results (Atkinson et al., 1998). These multi-level results indicate the presence of vertical wind shear in the equatorial region.

Above the cloud top, photochemistry-produced aerosols, e.g., $PH_3$, $NH_3$ from the troposphere (~100 – 10000 mbar) and hydrocarbons from the stratosphere (~0.001 – 100 mbar), are commonly referred to as "hazes" (West et al., 2004; Fletcher et al., 2020). The opacity and albedo of hazes vary, and their features appear less contrasted than clouds. Haze influences the sensitivity of thermal infrared observations used to detect the stratosphere. This poses challenges for accurate measurements of zonal winds at higher altitudes (Fry & Sromovsky, 2023).

Most recently, Hueso et al. (2023) analyzed images captured by the Near Infrared Camera (NIRCam, Horner & Rieke, 2004) on the James Webb Space Telescope (JWST) to obtain the zonal winds at various latitudes and altitudes. They compared 10-hr separated F164N images ($CH_4$ absorption band at 1.64 μm), F212N images ($H_2 – H_2$ and $H_2$ – He collision-induced absorption band at 2.12 μm), and F335M images ($CH_4$ absorption band at 3.35 μm) (Tokunaga & Vacca, 2005). They found an intense, narrow atmospheric jet in the lower stratosphere at Jupiter's equator. The velocity in the jet increases with increasing altitude, suggesting that Jupiter's equatorial stratospheric oscillation (JESO, Antuñano et al., 2021) likely penetrates the upper troposphere. JESO has a variable period of 3.9 to 5.7 years and shares some common characteristics with the quasi-biennial oscillation (QBO) on Earth (Antuñano et al., 2020; Giles et al., 2020). Temperature variations in the upper troposphere at about 330 mbar at the equator have been found to anticorrelate with those in the stratosphere. Such a relationship indicates the downward propagation of temperature anomalies (Antuñano et al., 2021; Orton et al., 2022). A recent HST study in the 890-nm methane band reports mixed evidence for the propagation of the JESO into the upper troposphere (Sánchez-Arregui et al., 2025). Thus, more detailed research on the relationship between stratospheric and tropospheric dynamics is necessary.

Based on these previous studies, a relatively comprehensive view of Jupiter's equatorial dynamics extends from the cloud tops to the stratosphere. At the cloud top, zonal winds remain globally stable with some variations. The newly discovered jet, which strengthens with increasing altitude, suggests a strong coupling between the troposphere and the stratosphere. More detailed measurements at different altitudes across the troposphere may help understand the connection, especially in the near-infrared wavelength range, where hazes are observable between the cloud top and the stratosphere.

In this paper, we use Cassini/VIMS infrared spectral image data spanning 0.85 µm to 5.1 µm to study Jovian zonal wind variability at different altitudes. In Sections 2 and 3, we introduce the spectral data and the methodology. Section 4 presents the zonal wind profiles from different wavelength image data using our correlation method and shows the vertical structure of zonal winds at the equator. Compared to the previous measurement, we discuss the relationship between stratospheric oscillations and zonal winds. Conclusions are drawn in Section 5.

## 2 Cassini/VIMS Data Sets and Data Processing

The Cassini spacecraft reached Jupiter in October 2000 and began its nearly 5-month flyby mission. Cassini was equipped with the advanced Visual and Infrared Mapping Spectrometer (VIMS), which consists of two distinct mapping spectrometers for different wavelength ranges: VIMS-V from 0.35 µm to 1.0 µm and VIMS-IR from 0.85 µm to 5.1µm (Brown et al., 2003). Our study uses data from the VIMS-IR wavelength range, divided into 256 wavelength bands with a spectral sampling of 0.0166 µm. If not specified, the wavelength given below is the central wavelength of each 0.0166 µm-wide band. During its flyby, VIMS collected more than 2,000 cube images. The size of the cube images collected could be up to 64 × 64 pixels.

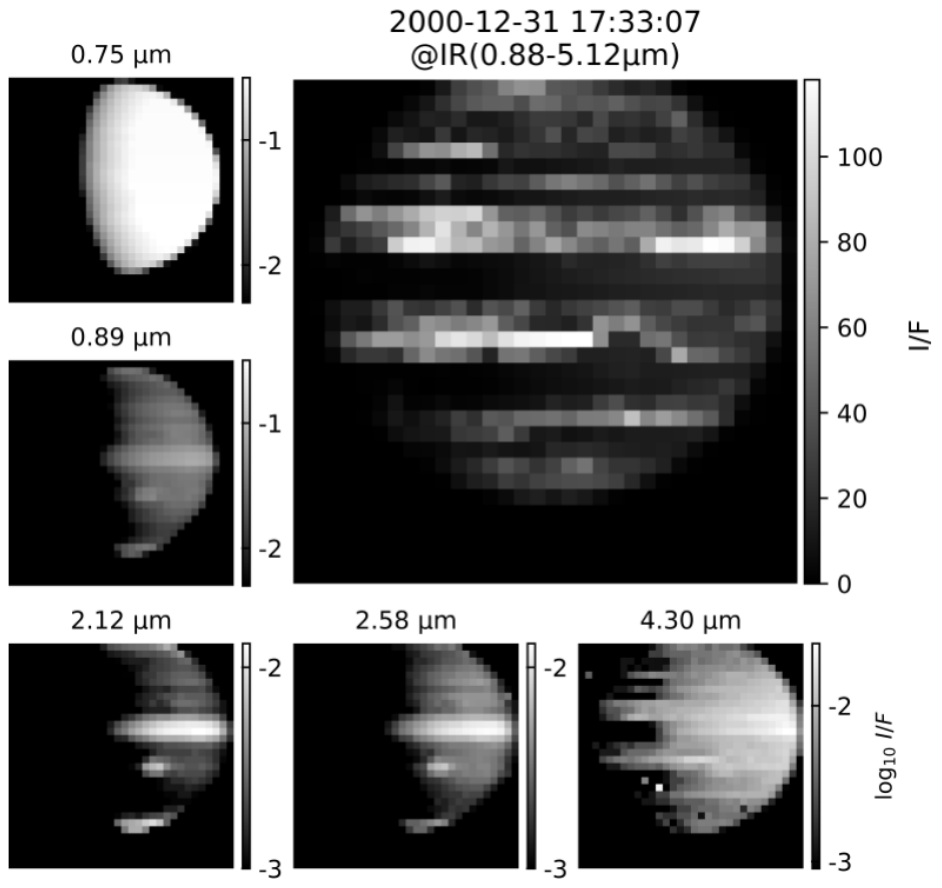

**Figure 1**. Cassini/VIMS images of Jupiter on 2000 December 31 crossing the entire wavelength range at five representative wavebands of 0.75 µm, 0.89 µm, 2.12 µm,

2.58 μm, and 4.31 μm. The brightness scale corresponds to I/F values as expressed in the color bar associated to each image.

The raw cube images are processed and calibrated using Integrated Software for Imagers and Spectrometers (ISIS, Anderson et al., 2004). ISIS relies on SPICE for calibration, including spacecraft ephemerides, instrument orientation, and mission event kernels to accurately map images onto Jupiter's surface (Mccord et al., 2004). After processing, we obtain radiometric and navigation information, including apparent reflectance (I/F), latitude, longitude, phase angle, incidence angle, and emission angle of every pixel in the cube images at each wavelength band. Detailed instructions are available at https://vims.univ-nantes.fr/info/isis-calibration. Figure 1 shows example cube images of I/F on 2000 December 31, spanning the entire VIMS-IR wavelength range and at five representative wavebands: 0.75 μm, 0.89 μm, 2.12 μm, 2.56 μm, and 4.3 μm.

To perform our study, we look for cube images taken over a sufficiently long and continuous period. By searching all 2000+ images, we chose the images taken on 2001 January 15, 9:42 UT – January 16, 3:22 UT. The period is the longest one with near continuous data in the 5-month flyby phase. However, the VIMS only covered a small part of the Jupiter disk during this period. An example cube image from this period is shown in Figure 2. According to the naming conventions of the cube files, the horizontal axis of the image is called 'samples', and the vertical axis is called 'lines'. The tilt of Jupiter's disk relative to the viewer's perspective in the image is ~3°. As a result, the measured wind speed represents only the component of the true wind speed projected by the 3° tilt. However, this effect is so small that it can be considered negligible. This allows us to assume that pixels on the same line are situated within the same latitude range. For example, three pixels at the 7th line, denoted by three color-coded dots, have center positions between 1°S and 1°N. When considering the full pixel latitudinal extent, their coverage extends from 4°S to 4°N. Thus, these three pixels represent the equatorial zone (EZ). The pixel bands shown in Figure 2 are bounded by latitude lines at 58°N, 46°N, 36°N, 27.4°N, 19.3°N, 11.5°N, 3.8°N, 3.6°S, 11°S, 18°S, 26.6°S, 34°S, and 44°S. Since Jupiter has a fast rotation, the spectral data of each pixel varies with time, which provides us with the information of the zonal winds at different latitudes.

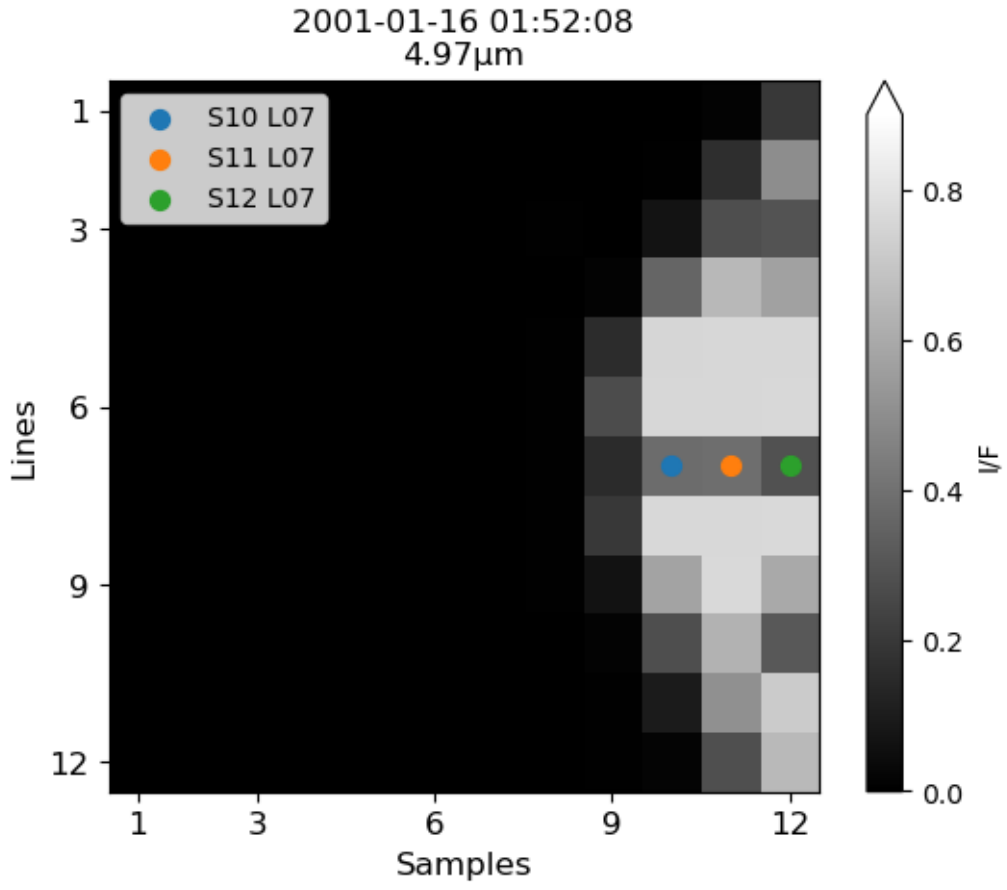

**Figure 2.** An example cube image from our interest period at 4.97 μm. Jupiter's outline is visible. Three pixels, denoted by three color-coded dots, have center positions at 0.9°N, 86.5°E (S10 L07), 0.2°N, 72.9°E (S11 L07), and 0.3°S, 61.1°E (S12 L07).

## 3 Methodology of Zonal Wind Measurements

### 3.1 A Correlation-Based Method

Correlation-based methods are widely used to obtain Jupiter's zonal winds. One method is to compare zonal scans of the planet from two time-separated images to determine the longitudinal displacement that best matches the features present in the zonal scans. According to the time difference between the two images, one can calculate the velocity of the target latitude (Limaye, 1989; García-Melendo, 2001; Porco et al., 2003; Asay-Davis et al., 2011; Tollefson et al., 2017; Johnson et al., 2018; Hueso et al., 2023; Sánchez-Arregui et al., 2025). Another method is to track several long-lived local features over a long time without the use of correlation techniques (Li et al., 2006; Sánchez-Lavega et al., 2008). The east-west components of the velocity fields are then averaged to obtain the average velocities in a specific latitude box. Besides, zonal winds can be obtained by observing Doppler shifts induced by the winds on specific spectral lines at the Jovian limb (Cavalié et al., 2021) or indirectly inferred from temperature measurements by applying the thermal wind equation (Flasar et al., 2004; Benmahi et al., 2021).

VIMS data have low spatial resolutions and a non-uniform cadence. Thus, they were never used to study the Jovian zonal winds. However, VIMS's broad wavelength coverage is a strength and motivates us to develop a correlation-based method to infer zonal winds from spectral images. Figure 3 shows the time evolution of the I/F values for the three pixels (denoted by three color-coded dots in Figure 2) at the equator. As Jupiter rotates and the Cassini viewing geometry gradually changes, each fixed pixel samples a sequence of longitudes at nearly constant latitude. The resulting I/F value variations therefore correspond to a zonal scan of the planet driven by Jupiter's rotation. Since the three pixels are located at nearly the same latitude, they scan similar features, allowing for redundancy and a better control of errors. The data cadence ranges from 117 s to 838 s, and due to overexposure and other instrumental issues, some data gaps occur, such as around 16:00 UT on January 15. All these issues have to be taken into account.

First, the big data gap around 16:00 UT on January 15 allows us to separate the data into two windows: the base window, shown in the red rectangle on the left in Figure 3, and the target window, shown in the blue rectangle on the right. The base window contains 127 data points per pixel and spans about 6 hr, from 9:42 UT to 15:24 UT on 2001 January 15. Thus, light curves for these pixels over this base window cover 60% of a full rotation of the planet. The target window contains 239 data points for each pixel and spans about 11 hr from 2001 January 15, 16:35 UT to January 16, 3:22 UT, covering more than a Jovian rotation period. We then search the target window for the segment that best matches the data profile in the base window by calculating the correlation coefficient (CC). The CC value of greater than 0.8 is considered significant. However, if the resultant CC profile contains multiple similar peaks without a clear, prominent one (as illustrated in Figure 4b), the search has failed and is therefore discarded. For those successful matches, the time shift between the base window and the best-match segment in the target window (i.e., the one with the highest CC value) is then converted to the velocity.

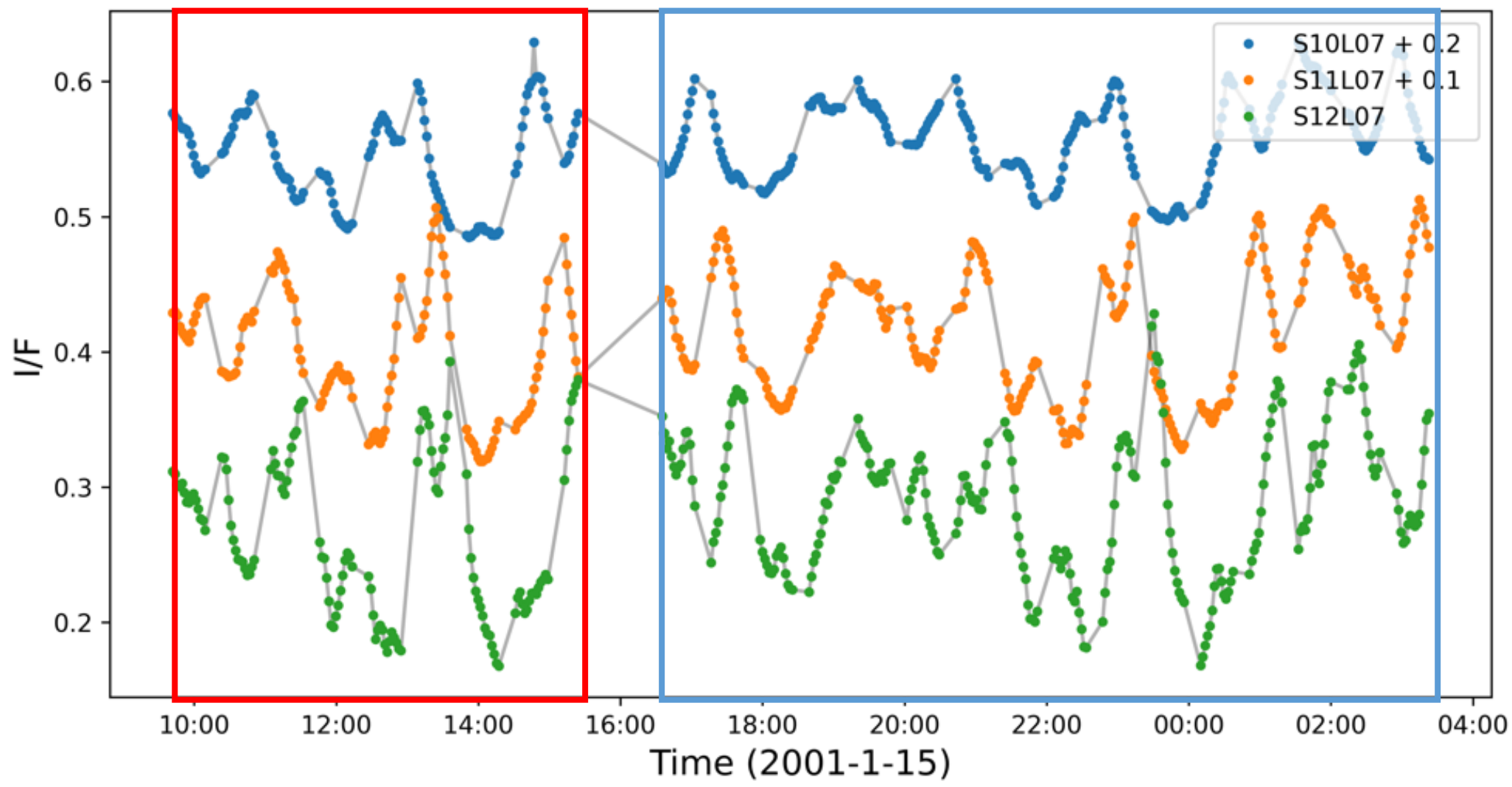

**Figure 3.** The I/F profiles at 4.97 μm for the three selected pixels in the equatorial region of Jupiter are indicated in Figure 2. To avoid a mixture, two of the three profiles are shifted upward by 0.1 and 0.2, respectively. The colored dots represent original non-uniformly sampled data points. The solid lines represent profiles formed by connecting these points for comparison. The red rectangle indicates the base window, and the blue rectangle is the target window.

Second, since the data were sampled at a non-uniform cadence, we uniformly resampled the data using linear interpolation before searching for the highest CC value. The resample is applied to both windows to make the processed data points uniformly separated by 117 s. We choose the 117s interval because it corresponds to the largest time differences between images in our dataset, allowing the interpolated data to closely follow the original measurements while preserving as many details as possible. The base window has 127 original non-uniformly sampled data points. Every one of the 127 original data points is chosen as the anchor point for the uniform resampling, leading to a total of 127 sets of uniformly distributed data points in the base window. Similarly, we may generate 239 sets of uniformly distributed data points in the target window. For each data set in the base window, we search for the best-matching segment in each data set in the target window by computing the CC values over discrete time shifts. For each pair of base and target windows, the shift corresponding to the highest CC value is converted into a suggested velocity. This procedure generates 127 x 239 = 30353 suggested velocities for each pixel.

In principle, we can get 30353 x 3 suggested velocities for the three profiles shown in Figure 3. However, as shown in Figure 4a, the CC profiles of three equatorial pixels show that peak positions may not be consistent. The green one is on the left of the other two and obviously lower. Such an inconsistency at the same latitude could be attributed to the low cadence, low spatial resolution, etc. Here, we retain the velocity corresponding to the highest peak (e.g., the blue CC peak), and we additionally retain other peaks only if their CC values differ from the maximum by no more than 0.1 lower (e.g., the orange CC peak in this case). As a result, we may finally obtain one to three sets of 30353 velocities for a latitudinal band, which are then put together to obtain the distribution of the derived velocities as shown in Figure 5a.

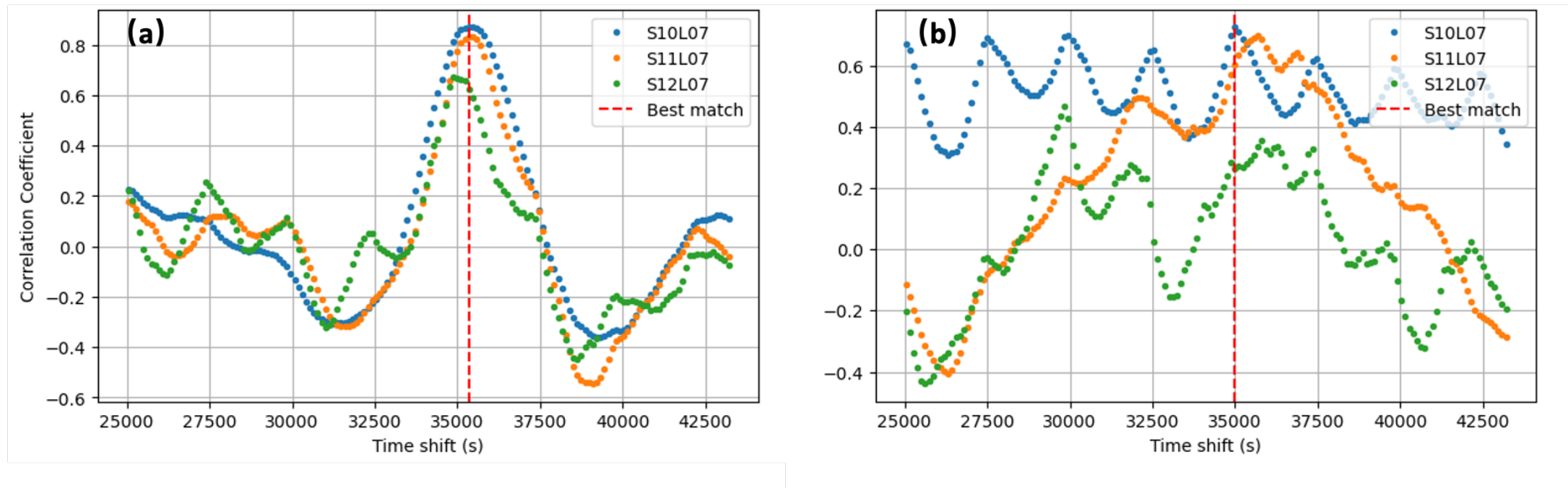

**Figure 4.** (a) A successful case of CC profiles for the three selected pixels. The red dashed vertical line indicates the maximum CC peak. (b) A failed case for comparison.

From Figure 5, we find that the derived velocities span a notable range due to multiple sources of systematic uncertainties, e.g., the low and non-uniform cadence of the data, the morphological evolution of atmospheric features, wave-like disturbances, and so on. But they roughly follow a single- or double-Gaussian distribution. Double-Gaussian or flattened single-Gaussian (i.e., the full width at half maximum is larger than 100 m/s) distributions, as illustrated in Figure 5b and c, mean that the error is too large to obtain a reliable velocity. Thus, we only retain the cases that exhibit a single and narrow Gaussian distribution. Then, we use a Gaussian function to fit the distribution and derive the most trusted velocity at the Gaussian peak. The full width at half maximum (FWHM) is adopted as the uncertainty. To mitigate the influence of unexpected extreme velocities, such as those exceeding 350 m/s in Figure 5, data points beyond the 90th percentile are discarded from the fitting.

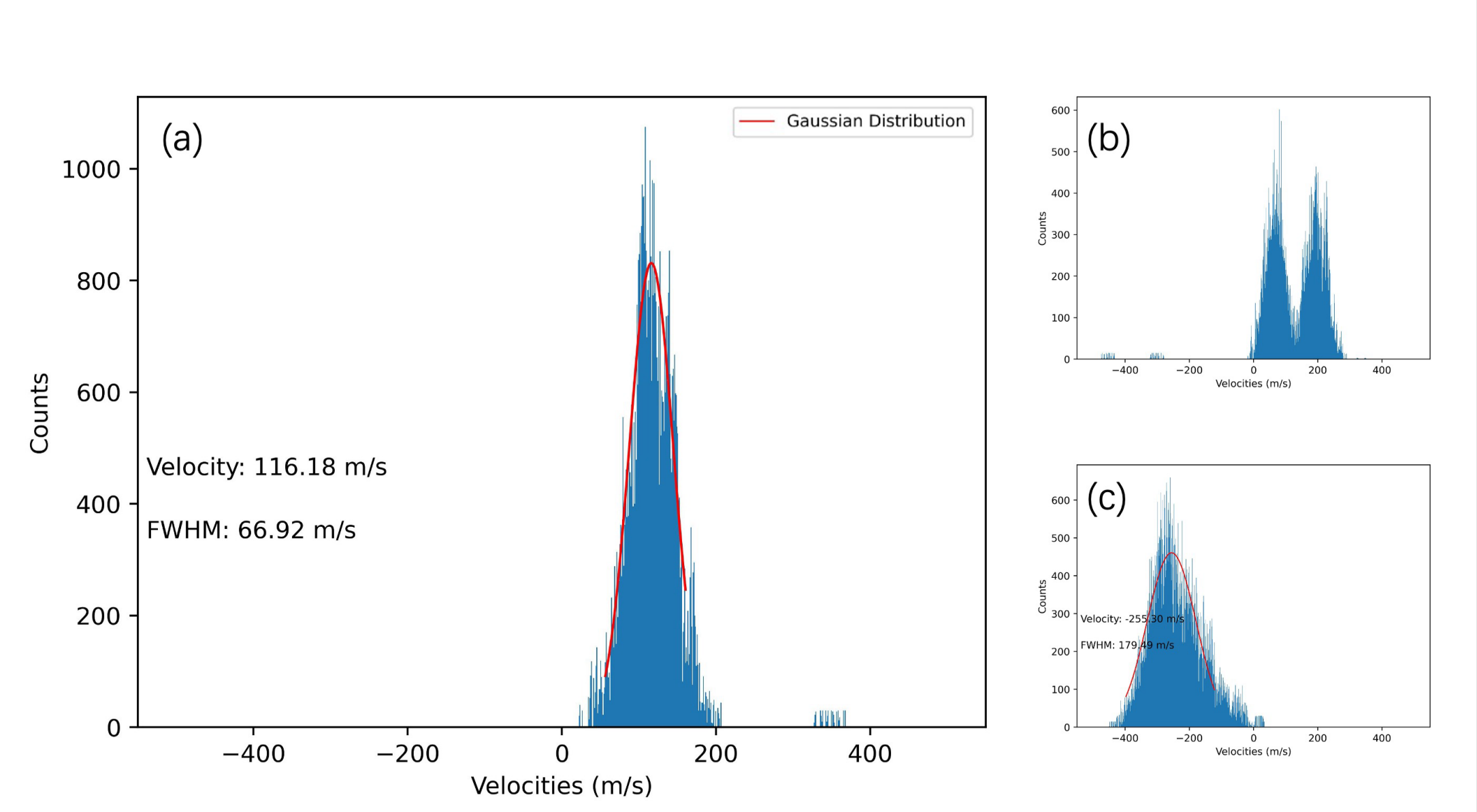

**Figure 5.** Distribution of the velocities derived from the correlation procedure. The red curve shows the Gaussian fitting result. (a) A successful case showing a single and narrow Gaussian distribution. (b) A failed case showing a double Gaussian distribution. (c) A failed case with a large FWHM, suggesting high uncertainty.

## 3.2 Velocity Calibration

Although the above method provides an estimate of the wind velocity and its uncertainty, the result may still deviate from the actual zonal wind velocity due to the VIMS data's small navigation errors combined with the low spatial resolution of the images lead to a large uncertainty in the longitudinal position of each pixel of about 10°, corresponding to a possible deviation of 350 m/s in the derived velocity. Though the resulting uncertainty in wind velocity appears large for a single pixel, it becomes a systematic error when averaged across a series of pixels. Thus, we may reduce this error by generating a velocity calibration curve obtained through the following experiments.

We first construct the I/F profile along the longitude of a certain latitude band based on the observed I/F values in the target window. In other words, we convert the time series of I/F to the longitudinal distribution. Since the time series of I/F is non-uniformly distributed, the constructed longitudinal distribution also has a non-uniform spatial separation varying from about 0.1° to 1.5°. Then, we increase its spatial separation to a uniform value of 0.1° using linear interpolation, which corresponds to a velocity uncertainty of less than 5 m/s. The constructed I/F profile retains the basic features of the atmospheric variations during the period of interest. Then we let this longitudinal profile of I/F rotate at a given velocity to simulate zonal motion, and use the same cadence and spatial resolution as the VIMS observations to sample the I/F values. Due to the non-uniform cadence of the data, we sample 239 sets of I/F profiles at the times of each data point in the target window. Through these steps, we obtain 239 time series of I/F in the base and target windows, with the real velocity known and navigation errors included. Then we derive the wind velocity from the constructed data through our correlation-based method introduced in the last subsection.

As expected, the derived velocity deviates from the true velocity due to the low cadence, non-uniformity, and low  spatial resolution of the data points. Since we generate 239 sets of constructed I/F profiles for a given real velocity, we may obtain 239 derived velocities. Further, we test the procedure for values of velocities from -100 m/s to 200 m/s in steps of 1 m/s to establish a relationship between the derived and real velocities. Figure 6 shows the result of these experiments. The orange line is the calibration curve showing the median of the 239 derived velocities versus the real velocity, and the blue shadow shows the calibration curve's uncertainty, which is the median of the FWHM values.

The calibration curve and its associated uncertainty, derived for each latitude band, are applied to the velocities derived from the observations. For example, if we derive a velocity of 100 m/s, we may draw a horizontal line at 100 m/s, as shown in Figure 6. It will yield three intersections with the orange calibration curve and the blue uncertainty region. The intersection with the calibration curve yields the true velocity, and the other intersections yield the uncertainties. In this example, a derived velocity of 100 m/s corresponds to a calibrated velocity of 65 m/s, with an uncertainty range from 38 m/s to 97 m/s. The final calibrated velocities and uncertainties are adopted as the zonal winds.

In summary, the low temporal and spatial resolution and the non-uniform sample rate of the Cassini/VIMS data result in large errors in estimating the zonal wind velocity. For the errors from the low and non-uniform cadence, the procedures in Sec.3.1 have been carefully treated. For the low spatial resolution, which will lead to position errors or navigation errors, the experiment in Sec. 3.2 is particularly designed to generate the velocity calibration curve to further correct the derived velocity. Besides the above errors, there are additional errors due to morphological evolution of atmospheric features. These are actually considered in the Gaussian fitting of the 30353 derived velocities in Sec. 3.1.

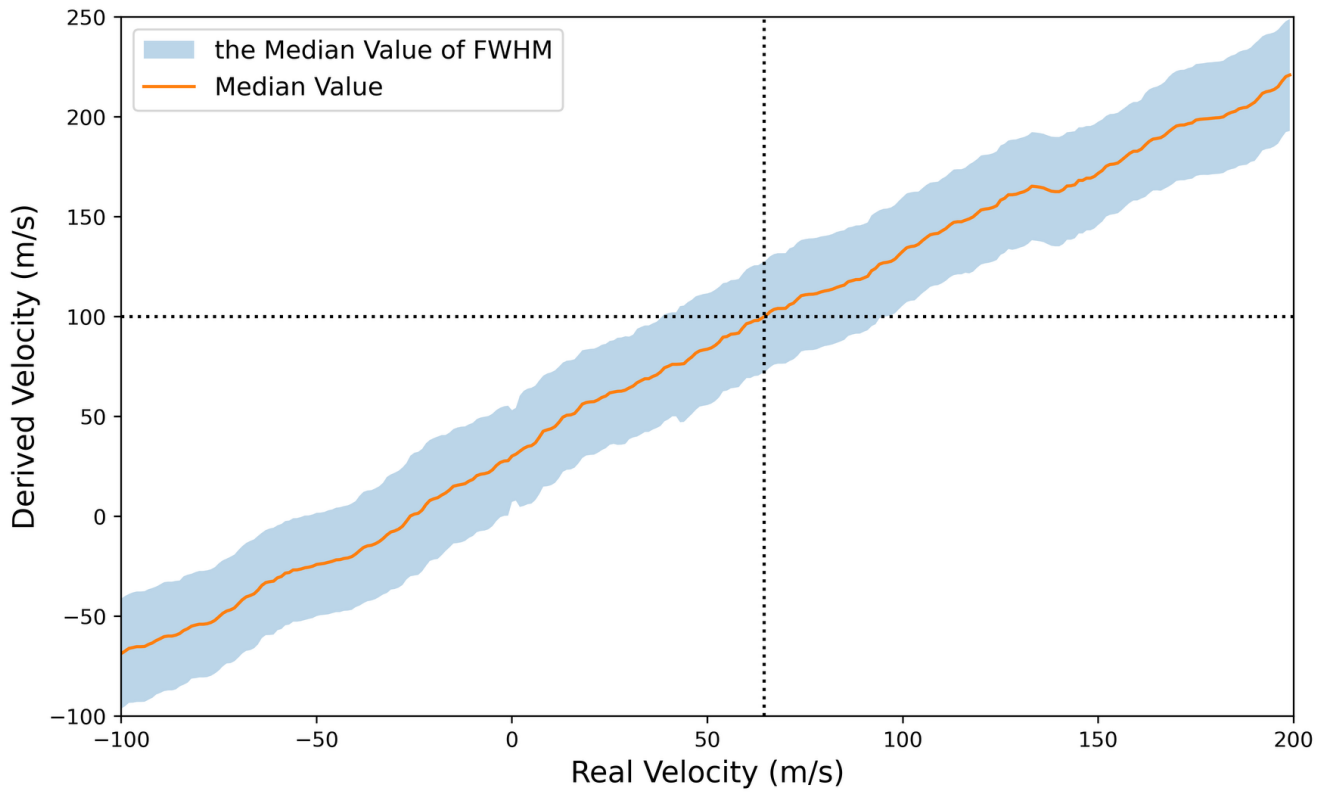

**Figure 6.** Relationship between the actual velocity and the real velocity. The orange solid line gives the median value of 239 derived velocities. The blue-shaded area was constructed using the median of the corresponding FWHM values.

## 4 Results and Discussion

Cassini/VIMS-IR covers a broad wavelength range with 256 bands. The spectrum, shown as the I/F value over the visible disk, is given by the red line in Figure 7. The near-infrared spectrum is primarily composed of sunlight reflected by Jupiter, with significant absorption bands resulting from gases such as methane ($CH_4$), phosphine ($PH_3$), and ammonia ($NH_3$). Here, we focus on three narrow wavebands as indicated by the color-coded vertical stripes in Figure 7. Two of them are centered at 2.58 µm and 4.31 µm, corresponding to $CH_4$ and $PH_3$ absorption bands, respectively, and the waveband widths are 0.048 and 0.032 µm, respectively (Baines et al. 2005, Malathy Devi et al. 2014, Lemière et al. 2021). The other waveband, centered at 2.12 µm with a bandwidth of 0.032 µm, is selected for comparison because it overlaps with the narrowband filter F212N from JWST/NIRCam studied by Hueso et al. (2023). In Hueso et al. (2023), the filter F164N covering 1.6 - 1.7 µm and F335M covering 3.19 – 3.54 µm were also investigated. However, we do not include an analysis of these wavelengths here because we did not observe significant features in the images at these wavebands, which makes it difficult to search for the highest CC value. In addition, the broad wavelength range of F335M filter may lead to tracking features at different altitudes, where wind shear can affect the measurements.

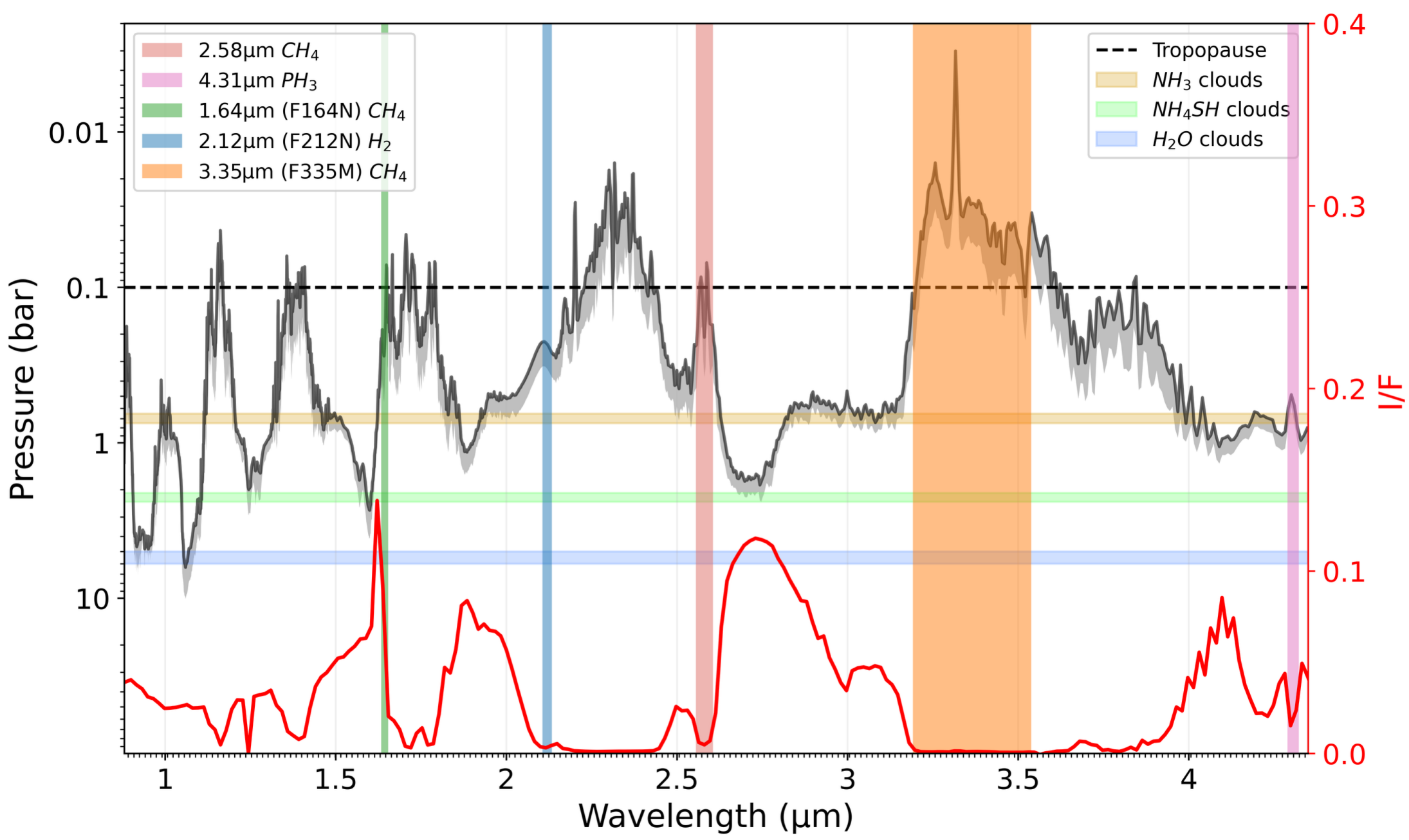

**Figure 7.** The pressure-wavelength response curve based on Sromovsky & Fry (2018). The black line represents the response curve with the two-way vertical optical depth (τ) equal to 1, and the shaded region gives the range where τ varies from 1 to 2 (Hueso et al. 2023). Key absorption features for different molecules are highlighted with vertical color columns: 2.58 µm $CH_4$ in brown, 4.31 µm $PH_3$ in pink, 2.12 µm

(F212N) $H_2$ in blue, 1.64 µm (F164N) $CH_4$ in green, and 3.35 µm (F335M) $CH_4$ in orange. Horizontal color bars indicate the layers of various clouds: $NH_3$-clouds in beige at the top of the cloud layer, $NH_4SH$-clouds in green, and $H_2O$-clouds in blue. The horizontal dashed line represents the tropopause. The red curve shows the I/F values corresponding to different wavebands, scaled by the y-axis on the right.

We use the Sromovsky & Fry (2018) model to establish the correspondence between the wavelengths and the pressure, as shown by the shadowed black line in Figure 7. The model assumes the absence of aerosols in Jupiter's atmosphere and that incoming solar radiation is dissipated only through gas absorption or Rayleigh scattering. The actual correspondence still depends on cloud and haze distributions and varies across different regions of Jupiter (Sromovsky & Fry, 2010, 2018). Based on Figure 7, we adopted an optical depth range similar to that used by Hueso et al. (2023), with $\tau \approx 1$–2 for the model calculations. The 4.31 µm waveband primarily senses pressures around 517 – 638 mbar, close to the cloud top. The cube image at 2.58 µm exhibits lower contrast than that at 2.12 µm, indicating a large optical depth, roughly within the 273 – 433 mbar range. In contrast, the 2.12 µm waveband is optically thinner and senses altitude near 228 – 327 mbar. Overall, both the 2.12 µm and 2.58 µm wavebands probe atmospheric levels well above the cloud top.

## 4.1 Zonal Wind Profile

Figure 8 presents the zonal wind profiles from the three selected wavebands with our correlation-based method. Our analysis does not cover the entire latitude range for several reasons. First, the limited field of view did not cover high-latitude regions during the period of interest. Second, in areas near the Jovian limb, the highly inclined viewing angle makes it difficult to detect a prominent CC peak during correlation, limiting the measurements. Third, atmospheric activity in some regions may influence the correlation process, leading the retrieved motions to represent the motions of large-scale features, waves or other disturbances rather than true cloud and haze motions. Therefore, we focused on five latitude bands between 20°S and 20°N, each containing three available pixels.

Our derived zonal winds from the three wavebands are consistent with one another. They are also close to the cloud-top winds (the black lines) retrieved by Porco et al. (2003) based on Cassini/ISS images and/or the zonal wind profiles (the blue line) from JWST/NIRCam by Hueso et al. (2023), whose mean values within the left-hand latitudinal extents are shown as unfilled diamonds in Figure 8. However, at approximately 8°N and 8°S, corresponding to the northern and southern boundaries of the EZ, the derived wind velocities from adjacent levels show noticeable deviations of about 50 - 80 m/s. These mainly arise from comparisons between the 4.31 µm band (the purple circles) and the cloud top (the black diamonds), and between the 2.12 µm band (the blue circles) and F212N (the blue diamonds). These differences are likely due to atmospheric variability near the boundaries, where the hazes exhibit strong temporal variability, making it difficult to track consistent features. In addition, the presence of slow tracers in these regions may contribute to lower velocity values (West et al., 2004; Wong et al., 2008; García-Melendo et al., 2011a; Hueso et al., 2023).

We also observe deviations in the EZ. Figure 8 shows the zonal wind profile from the 4.31 μm waveband corresponding to 517 – 638 mbar close to the cloud top. Our derived velocity in the EZ is about 96 m/s. The derived velocity is close to the cloud-top velocity by Porco et al. (2003) and our uncertainty covers their value. This waveband contains the $PH_3$ absorption line which probes levels just above the cloud. Thus, the velocity deviation likely reflects the different motions of the $PH_3$ hazes. Figure 8 also shows the zonal wind profile derived from the 2.58 μm waveband, corresponding to a pressure level of 273–433 mbar. In the EZ, the derived velocity exceeds 109 m/s, exceeding the cloud-top velocity even when accounting for uncertainty. The result of $CH_4$'s sensitivity to this band represents a strong eastward jet of the $CH_4$ hazes at this level. For the 2.12 μm waveband, which probes altitude slightly above that of 2.58 μm, the derived EZ velocity is about 111 m/s, roughly 20 m/s higher than the mean value in Hueso et al. (2023). This deviation is likely due to the broader waveband used in this study compared to JWST, which makes it sensitive to a wider vertical range of the atmosphere. Despite the difference, the EZ velocity reported by Hueso et al. (2023) is a mean value; they obtained a wide range of velocities, from 60 to 150 m/s, due to the wind shear. Thus, our derived EZ velocity is consistent with their result.

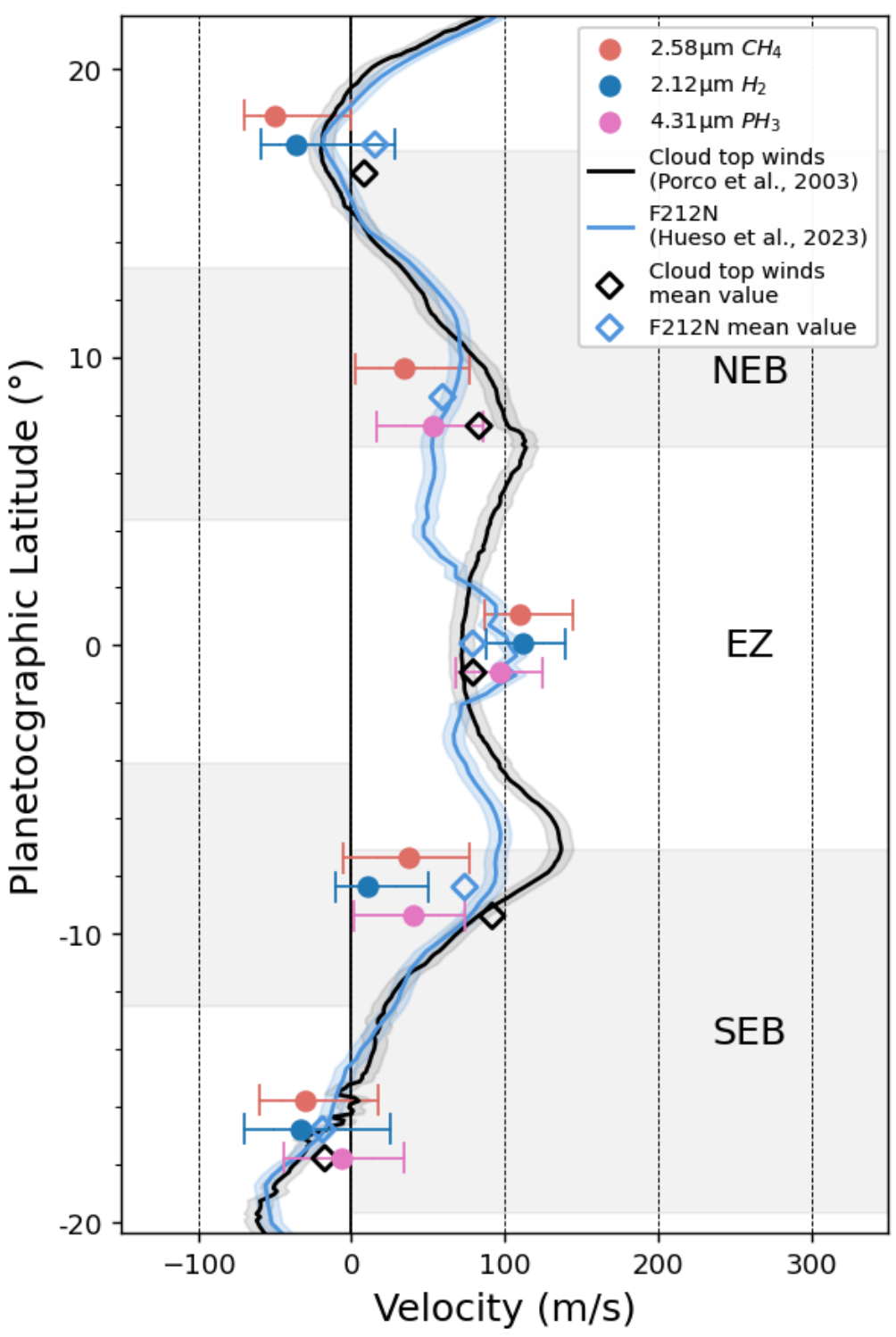

**Figure 8.** Zonal winds at various planetographic latitudes. The alternating white/gray stripes on the left-hand side of the vertical zero line indicate the latitudinal extents of the pixels of the images of interest, and those on the right-hand side indicate the locations of the belts and zones structure, with their names labeled on the right.

Filled circles and error bars represent the derived and calibrated wind velocities and uncertainties. For visualization, the circles are plotted with small latitude offsets to avoid overlap. The black and blue lines show the zonal wind profiles from the Cassini/ISS observations (Porco et al., 2003) and JWST/NIRCam observations (Hueso et al., 2023), respectively. Their shaded regions represent the mean of all standard deviations (1σ) between 20°S and 20°N, which is about 8 m/s. The unfilled diamonds represent the mean velocities over the latitudinal extents denoted by the stripes on the left-hand side. The positive value of velocity means an eastward zonal wind. Velocities in some regions are missing because the correlation results did not meet our criteria (i.e., the velocity distributions did not conform to a single, narrow Gaussian profile) and were therefore excluded.

## 4.2 Vertical Structure of Zonal Winds in the EZ

As shown in Figure 8, only the pixels at the equator are entirely located within the latitudinal extent of the EZ, and other pixels cross the boundaries of belts and zones. Therefore, we focus our analysis of the zonal winds on the EZ, where the spatial resolution is relatively higher than at other latitudes. Figure 9 shows the vertical structure of the zonal winds in the EZ. The filled circles represent the zonal winds derived from our measurements, while the unfilled circles indicate values from previous studies. The error bars indicate uncertainties in both pressure and velocity.

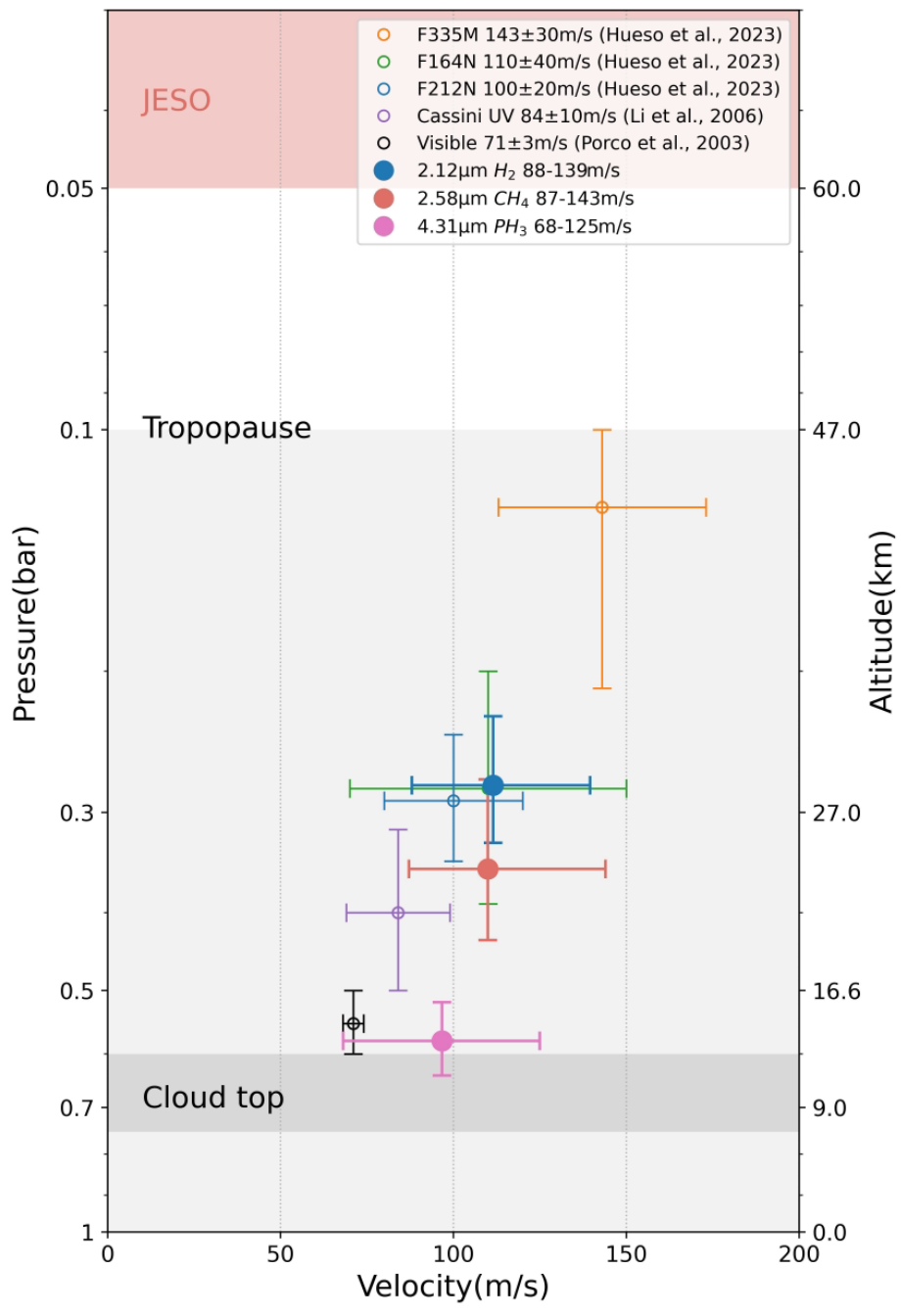

**Figure 9.** Vertical structure of the zonal winds in the EZ. Filled circles and their error bars represent the zonal winds and associated uncertainties derived from the three wavebands in this study. The vertical (pressure) error bars are obtained from the radiative transfer model presented in Figure 7. In contrast, the horizontal (velocity) error bars are based on the calibration described in Sec.3. Unfilled circles and their error bars represent zonal winds from previous measurements. Their vertical (pressure) uncertainties are taken from West et al. (2004) and Hueso et al. (2023),

whereas the horizontal (velocity) uncertainties correspond to the mean of all standard deviations (1σ) between 3°S and 3°N. The background is color-coded to distinguish between the troposphere and stratosphere, with the approximate location of the cloud top and the JESO marked.

It can be seen from Figure 9 that the velocity increases with altitude, from near the cloud top, where it is about 96 m/s, to the upper troposphere at 273 – 433 mbar, where it reaches about 111 m/s. This increase in zonal winds with altitude confirms the vertical wind shear in the troposphere. Although our derived velocities are larger than those at the same pressure levels, the wind shear looks weaker than previous results combined from Porco et al. (2003), Li et al. (2006), and Hueso et al. (2023); if we do not consider the uncertainties. One possible explanation is that the three studies and ours are based on data collected at four different dates over more than 20 years. Thus, variations in zonal winds on timescales ranging from one month to years, typically between 10 and 20 m/s (García-Melendo, 2001; Asay-Davis et al., 2011; Barrado-Izagirre et al., 2013), should be considered when discussing the differences between the results. Another possible explanation is that zonal winds are constant over the pressure range measured and that differences correspond to the combination of errors of each different method including the coarse latitudinal resolution of the Cassini/VIMS data combined with slight altitude variations from the different wavelengths used.

4.3 Mechanisms Driving Wind Shear Variability

One major cause of the variations is the equatorial stratospheric oscillations with a period of 3.9 to 5.7 years, as indicated by the shaded red area in Figure 9. According to the thermal wind relationship, the equatorial temperature anomaly is in thermal wind balance with the vertical shear of zonal wind. At the equator, the warm anomaly is associated with eastward zonal wind shear, while the cold anomaly corresponds to westward zonal wind shear (Tanii & Hasebe, 2002; Pascoe et al., 2005; Antuñano et al., 2020; Giles et al., 2020; Zhang et al., 2024). Temperature oscillations were detected at the equator in the stratosphere from 1983 to 2019 (Orton et al., 2023), suggesting a weak warm anomaly during our study period, consistent with weak eastward zonal wind shear. The persistent oscillations exert long-term variations on the background wind field.

In addition to stratospheric oscillations, atmospheric waves, e.g., small-scale and planetary-scale waves, with periods ranging from hours to days, can modulate the zonal wind and introduce transient effects on the observed wind shear. The specific impact of these waves depends on their characteristics, particularly their vertical wavelengths (Allison, 1990; Li et al., 2006; Simon-Miller & Gierasch, 2010; García-Melendo et al., 2011). However, these waves are superimposed on the background zonal wind, making it challenging to distinguish them.

In summary, differences in wind shear across studies can be driven by multiple factors. While stratospheric oscillations contribute to long-term variations and various waves introduce transient changes, the low spatial resolution remains a key limitation. Given the pixel scale, haze features smaller than the spatial resolution may lead to blending of I/F values across different altitudes sensed by the different wavelength bands. This blending impairs the ability to isolate individual tracer movements, making it difficult to distinguish between zonal winds and vertical variations caused by convective activity or rapidly evolving features at different altitudes. These small-scale structures may decrease in size, merge, or evolve over time, further complicating the interpretation of zonal wind measurements. Despite efforts to mitigate these limitations using the calibration curve, the constraints imposed by spatial and temporal resolution continue to make it challenging to separate these factors fully.

## 5 Conclusions

Over the past few decades, significant advances have been made in understanding Jupiter's zonal winds, primarily through multi-wavelength imaging. Previous studies have utilized various methods to derive zonal wind profiles, revealing a strong coupling between stratospheric oscillations and the troposphere. However, these studies still lack observations in the near-infrared wavelength bands for additional periods, which are crucial for investigating temporal variations in Jupiter's atmospheric dynamics. Given the scarcity of near-infrared wavelength images of Jupiter from Cassini/VIMS-IR, our study developed a new correlation-based method to investigate zonal winds.

We obtained the zonal winds at low latitudes for three wavebands. They are consistent with previous studies, though deviations suggest the zonal winds' complexity over altitude and time. Our analysis of the vertical structure of the zonal winds in the EZ revealed the eastward wind shear in the troposphere. The wind shear is weaker than the profile from the combination of previous studies, but the velocities are all larger if uncertainties are not considered. The differences may reflect factors influencing the zonal winds: stratospheric oscillations cause long-term variations, while various waves cause short-term variations. Future work should focus on identifying and characterizing these waves, particularly in the equatorial region, to refine our understanding of zonal wind variability.

Despite these challenges, our study utilizes the broad wavelength range of Cassini/VIMS, further improving the dataset's usefulness and offering new insights into Jupiter’s zonal winds. Additionally, our method provides an alternative to traditional cloud-tracking techniques, allowing analysis of lower-quality time-lapse images. In the future, we aim to explore potential wave phenomena in Jupiter’s atmosphere during the observation period, which will be vital for deepening our understanding of the planet’s atmospheric dynamics.

Time-series of multi-spectral data allow for wind retrievals at multiple heights. Similar techniques are currently being used to explore the rotation of brown dwarfs with multi-spectral time- series obtained by the James Webb Space Telescope (e.g. Biller et al., 2024). Future observations of Jupiter with the MAJIS multi-spectral instrument (Poulet et al., 2024) onboard the Jupiter Icy Moons Explorer (JUICE) during the approach phase to the planet in 2031 might consider the use of this technique to provide early wind data at multiple heights. Later observations from apojoves during the Jupiter tour (Fletcher et al., 2023) could take advantage of this technique to obtain regular winds at a low observational cost.

## Acknowledgments

We are grateful to all the reviewers for their constructive suggestions on my work. We thank the Cassini team for their invaluable work in planning and carrying out the mission and the ISIS software team for data calibration. The Cassini/VIMS data are available from NASA's Planetary Data System (https://pds-imaging.jpl.nasa.gov/volumes/vims.html). The work is supported by the National Key R&D Program of China (2025YFF0512400), the National Natural Science Foundation of China (42130203, and 42174213), and the Quantum Science and Technology-National Science and Technology Major Project (2021ZD0300302). We acknowledge the support from the National Space Science Data Center, National Science & Technology Infrastructure of China (www.nssdc.ac.cn).

## Open Research

The Cassini/VIMS data are available from NASA's Planetary Data System (https://pds-imaging.jpl.nasa.gov/volumes/vims.html). ISIS, a free and open-source software program, can be downloaded from Laura et al. (2023). The velocity calibration curves are stored in Ma (2025).

## Conflict of Interest Disclosure

The authors declare there are no conflicts of interest for this manuscript.